\documentclass{jetpl}
\usepackage{epsfig}

\twocolumn

\lat

\DeclareMathOperator{\Tr}{{\mathrm{Tr}}}
\DeclareMathOperator{\NP}{{\mathrm{NP}\!}}

\newcommand{\eq}[1]{(\ref{#1})}

\title{
\rightline{{\normalsize{ITEP-LAT/2007-10}} \vspace{5mm} }
Topological susceptibility in Yang-Mills theory \\ in the vacuum correlator method}

\rtitle{Topological susceptibility in Yang-Mills theory\ldots}

\sodtitle{Topological charge correlations in Yang-Mills theory in the field correlator method}

\author{M.\,N.\,Chernodub$^{*}$, I.\,E.\,Kozlov$^{*,**}$
}

\rauthor{Chernodub M.N., Kozlov I.E.}

\sodauthor{Chernodub, Kozlov}

\address{$^*$ Institute of Theoretical and Experimental Physics,
B.Cheremushkinskaya 25, 117218 Moscow, Russia\newline
$^{**}$ M.V. Lomonosov Moscow State University, Faculty of Physics, Moscow, 119992, Russia}

\abstract{We calculate the topological susceptibility of the
Yang-Mills vacuum using the field correlator method. Our estimate for the $SU(3)$ gauge group, $\chi^{1/4} = 196(7) \,\mbox{MeV}$,
is in a very good agreement with the results of recent numerical simulations of the Yang-Mills theory on the lattice.}


\begin{document}

\maketitle

The topology of the QCD vacuum and the hadron phenomenology are intimately related.
The celebrated example is given by the Witten-Veneziano formula~\cite{ref:Witten:Veneziano}
which provides a solution to the ``$U(1)$ problem'' by utilizing
the explicit breaking of the $U(1)$ axial symmetry in the large-$N$ QCD.
In a simplest case of $N_f$ massless quarks the formula relates the mass of the $\eta'$ meson,
\begin{eqnarray}
m^2_{\eta'} = \frac{2 N_f}{F^2_\pi} \, \chi\,,
\label{eq:Witten:Veneziano}
\end{eqnarray}
to the non-perturbative susceptibility
\begin{eqnarray}
\chi & = &
%\frac{1}{V}\langle\langle Q^2\rangle\rangle =
\lim_{V \to \infty}\frac{1}{V}\left\langle (Q - \langle Q\rangle)^2\right\rangle
%\nonumber\\ & = &
= \int\!\! {\mathrm d}^4\! z \, \langle q(0) q(z) \rangle,
\label{eq:Q:susc}
\end{eqnarray}
of the topological charge
\begin{eqnarray}
Q = \int {\mathrm d}^4 x\, q(x)\,,
\label{eq:top:charge}
\end{eqnarray}
calculated in the pure $SU(N)$ Yang-Mills theory, and to the pion decay constant $F_\pi$
(note that $\langle Q\rangle = 0$ due to $CP$--invariance of the Yang-Mills vacuum).
An exact value of the susceptibility $\chi$ is the subject of intensive numerical studies
in the lattice gauge theories~\cite{ref:interest:lattice,ref:chi:lattice}.

The local density $q(x)$ of the topological charge is
\begin{eqnarray}
q(x) = \frac{g^2}{32\pi^2} \epsilon_{\alpha\beta\mu\nu}  \Tr (F_{\alpha \beta}(x) F_{\mu\nu}(x))
\label{eq:Q}
\end{eqnarray}
where $F_{\mu\nu} = F_{\mu\nu}^a T^a$ is the non-Abelian field strength tensor
with the components
\begin{eqnarray}
F^a_{\mu\nu} = \partial_\mu A^a_\nu -  \partial_\nu A^a_\mu + g f^{abc} A^b_\mu A^c_\nu\,,
\label{eq:F}
\end{eqnarray}
and $T^a$, $a=1,\dots,N^2-1$ are the generators of the $SU(N)$ gauge group normalized by the
standard relation, $\Tr T^a T^b = \delta^{ab}/2$, and $f^{abc} = - 2 i \Tr [T^a,T^b] T^c$
is the totally asymmetric structure constant of the group.

In this paper we estimate the topological susceptibility~(\ref{eq:Q:susc}) in the formalism of gauge-invariant
non-local correlators~\cite{ref:correlators} which is reviewed in Ref.~\cite{ref:review,ref:review:two}. We work in Euclidean
space-time suitable for comparison of our results with the ones obtained in numerical simulations of
the lattice Yang-Mills theory.
A basic element of the field correlator method is a non-local quantity,
\begin{eqnarray}
G_{\mu\nu}(x;x_0) = \Phi(x_0;x) F_{\mu \nu}(x) \Phi(x;x_0)\,,
\label{eq:G}
\end{eqnarray}
where $\Phi(x;x_0) \equiv \Phi^\dagger (x_0,x)$ is an ordered exponent of the integral over
the gauge field $A_\mu = T^a A^a_\mu$ (the Schwinger line).
The integration is taken along the oriented path $\cC_{x,x_0}$ which stretches between
the points $x_0$ and $x$ of the space-time,
\begin{eqnarray}
\Phi(x;y) = \cP \exp\Bigl[ - i g \int\nolimits_x^y A_\mu (z) \, d z_\mu \Bigr]\,,
\label{eq:Phi}
\end{eqnarray}
where $\cP$ denotes path ordering.
Under the $SU(N)$ gauge transformation $\Omega$ the quantity~(\ref{eq:Phi})
changes as
\[
\Phi(x;y) \to \Omega(x) \Phi(x;y) \Omega^\dagger(y)\,.
\]
Due to this property the non-local object~(\ref{eq:G}) transforms
essentially locally,
\[
G_{\mu\nu}(x,x_0) \to \Omega(x_0) G_{\mu\nu}(x,x_0) \Omega^\dagger(x_0)\,,
\]
making it possible to construct various (non-local) gauge invariant objects of the general form
\begin{eqnarray}
D^{(n)}_{\mu_1 \dots \nu_n} \propto \Tr [G_{\mu_1\nu_1}(x_1;x_0) \dots G_{\mu_n\nu_n}(x_n;x_0)]\,.
\label{eq:Dn}
\end{eqnarray}
One may think about this quantity as of the $n$-point correlation function of the field strength
tensors~(\ref{eq:F}) covariantly shifted from the ``reference point'' $x_0$ to the points
$x_i$, $i=1,\dots,n$ along appropriate paths.
The simplest nontrivial quantity of the kind~(\ref{eq:Dn}) is the two-point correlator
(hereafter we omit the superscript $n=2$)
\begin{eqnarray}
D_{\mu\nu\alpha\beta}(x_1,x_2;x_0) = \frac{g^2}{N} \Tr [G_{\mu\nu}(x_1,x_0) \, G_{\alpha\beta}(x_2;x_0)]\,.
\label{eq:D2}
\end{eqnarray}
One may choose the paths connecting the points $x_i$, $i=0,1,2$ as straight lines and put the point
$x_0$ exactly on a straight line between the points $x_1$ and $x_2$.
Then the correlator~(\ref{eq:D2}) becomes a function of a single variable $z = x_1 - x_2$.

At zero temperature the correlator~(\ref{eq:D2}) can be parameterized by two
scalar functions $D$ and $D_1$
\begin{eqnarray}
&& D_{\mu\nu\alpha\beta}(z) = \Bigl(\delta_{\mu\alpha} \delta_{\nu\beta}
- \delta_{\mu\beta} \delta_{\nu\alpha} \Bigr) \, D(z^2)
\label{eq:g2D} \\
&& \hspace{7mm} + \frac{1}{2} \Bigl[
\frac{\partial}{\partial z_\mu} (z_\alpha \delta_{\nu\beta} - z_\beta \delta_{\nu\alpha}) -
(\mu \leftrightarrow \nu) \Bigr]\, D_1(z^2)\,.
\nonumber
\end{eqnarray}
Perturbatively, both functions are divergent as $1/|z|^4$ at short distances, $|z| \to 0$.
However, besides the perturbative part the structure functions possess also a
non-perturbative part which is relevant to the long--distance physics.
The lattice simulations~\cite{ref:Adriano:pure,ref:Adriano:QCD} indicate that
the structure functions can be well described as follows
\begin{eqnarray}
\begin{array}{rcl}
D(z^2)   & = & A_0 \, e^{ - |z|/T_g} + \frac{b_0}{|z|^4} \, e^{ - |z|/\lambda} \,, \\
D_1(z^2) & = & A_1 \, e^{ - |z|/T_g} + \frac{b_1}{|z|^4} \, e^{ - |z|/\lambda}  \,,
\end{array}
\label{eq:D:D1}
\end{eqnarray}
where the first terms in both expressions correspond to a non-perturbative contribution
while the last two terms contain the perturbative parts. The functions $D$ and $D_1$ are parameterized
by the correlation lengths $T_g$ and $\lambda$, and by the prefactors $A_i$ and $b_i$, $i=0,1$.

On general grounds one may expect that the correlators~(\ref{eq:Dn}) should
depend not only on the points $x_i$, but also on the auxiliary variables such as $x_0$ and shapes of
the paths $\cC_{x_n,x_0}$ entering the phase factors~(\ref{eq:Phi}) used in Eq.~(\ref{eq:G}). Perturbative
corrections coming from the Schwinger lines may provide an additional
contribution to the correlation lengths. However, a physical non-perturbative observable -- if expressed
in terms of the correlators~(\ref{eq:Dn}) only -- should be independent of both the shape of the Schwinger lines
and the ultraviolet cutoff.

In what follows we consider the quadratic correlator~(\ref{eq:D2}) for which various numerical estimations
are available~\cite{ref:Adriano:pure,ref:Adriano:QCD,ref:Michael:smoothing}. The numerical results
were obtained in lattice simulations with the use of the so-called cooling method which makes it possible
to get rid of short-range lattice artifacts while leaving the long-range physics intact.
The cooling removes ultraviolet suppression of long Schwinger lines $\Phi(x,y)$, which may originate from
perturbative fluctuations. As a result of the (soft) cooling, the correlation lengths $T_g$
and $\lambda$ in the correlators~(\ref{eq:D:D1}) do not depend on the ultraviolet cutoff both
in pure Yang-Mills theory~\cite{ref:Adriano:pure} and in QCD with dynamical quarks~\cite{ref:Adriano:pure}.
Similar results were obtained in Ref.~\cite{ref:Michael:smoothing} with the use of the so called renormalization
group smoothing which is an alternative to the cooling.

Another subtlety of the lattice results is the dependence of the correlators on the shape of the Schwinger line
connecting the points $x_0$, $x_1$ and $x_2$ in the definition of the correlator~(\ref{eq:D2}). It turns out that the
dependence of the prefactors $A_i$ and $b_i$, $i=0,1$ is rather strong, while the correlation lengths
$T_g$ and $\lambda$ are insensitive with respect to the variations of the shape of the line~\cite{ref:Adriano:shape}.
Thus our results should be understood with the prescription that the perturbative ultraviolet corrections
are subtracted and the Schwinger line itself is chosen to take a most natural shape forming
straight path stretched between the points $x_1$ and $x_2$ in the definition of the correlation function~(\ref{eq:D2}).
In this prescription the correlator is clearly independent of the reference point $x_0$ provided it is located
on the Schwinger line stretched between the points $x_1$ and $x_2$.

It is argued~\cite{ref:review} that in the Yang-Mills theory and in QCD the dominant contribution to various
observables is given by the lowest bilocal correlator. This observation is often call as the ``Gaussian dominance''.
The validity of the Gaussian approximation is supported, for example, by the argument, that
the contribution of the bilocal correlations to the interactions between
colored particles dominates the contribution coming from the higher-order correlators. The dominance could be
interpreted either in stochastic picture (any $n>2$ correlator contributes much less then the binomial one) or
in the coherent picture which postulates that there is a strong cancelation between all $n>2$ correlator so that
the result is given only by the binomial contribution~\cite{ref:Shevchenko:PRL}.
We take an advantage of the stochastic scenario assuming that all higher order correlators are
suppressed with respect to the leading Gaussian contribution.

In the Gaussian vacuum all $n$--point correlators of field strengths are assumed to be factorized into the bilocal
correlators. In particular, this means that all odd correlators are zero in the Gaussian approximation. The
factorization of even correlation functions goes according the scheme~\cite{ref:review}:
\begin{eqnarray}
& & \langle G^{a_1}(1) G^{a_2}(2) \dots G^{a_{2n}}(2n)\rangle \sim
(\delta^{a_1 a_2} \dots \delta^{a_{2n-1} a_{2n}}) \nonumber\\
& & \times \langle \Tr G(1) G(2) \rangle \dots \langle \Tr G(2n-1) G(2n) \rangle \\
& & + \mbox{all permutations}\,, \nonumber
\end{eqnarray}
where the color arguments of the field strength tensors are written in a short form.

The quadratic correlation function~(\ref{eq:Q:susc})
of the topological densities involves the correlator of the four field strengths operators located
in two points of the space time. A related four-point correlator of field strengths can be calculated in the
Gaussian approximation according to the factorization scheme described above:
\begin{eqnarray}
&   & \langle F^a_{\alpha \beta}(0)F^b_{\gamma \delta}(0)F^c_{\mu \nu}(z) F^d_{\rho \theta}(z) \rangle
\nonumber \\
& = & \langle F^a_{\alpha \beta}(0)F^b_{\gamma \delta}(0) \rangle \, \langle F^c_{\mu \nu}(z)F^d_{\rho \theta}(z) \rangle
\nonumber \\
& & + \langle F^a_{\alpha \beta}(0) F^c_{\mu \nu}(z) \rangle \, \langle F^b_{\gamma \delta}(0)F^d_{\rho \theta}(z) \rangle
\nonumber \\
& & + \langle F^a_{\alpha \beta}(0)F^d_{\rho \theta}(z) \rangle \, \langle F^b_{\gamma \delta}(0)F^c_{\mu \nu}(z) \rangle
\label{eq:four:point}\\
& \equiv &
\frac{4}{g^4}
\left(\frac{\mathrm N}{{\mathrm N}^2-1}\right)^2
\Bigl[
\delta^{a b} \delta^{c d} \, D_{\alpha \beta \gamma \delta}(0) D_{\mu \nu \rho \theta}(0) \nonumber\\
& & +
\delta^{a c} \delta^{b d}\, D_{\alpha \beta \mu \nu}(z)\, D_{\gamma \delta \rho \theta}(z) \nonumber\\
& & + \delta^{a d} \delta^{b c} \, D_{\alpha \beta \rho \theta}(z)\, D_{\gamma \delta \mu \nu}(z) \Bigr]\,.
\nonumber
\end{eqnarray}

In this calculation we identified the field strength tensors $F_{\mu\nu}$ -- used in the definition of the
topological charge density~(\ref{eq:Q}) -- with the covariantly transformed field strength tensors $G_{\mu\nu}$,
Eq.~(\ref{eq:G}), used in the definition of the gauge--invariant bilocal correlators~(\ref{eq:D2}).
We used the observation that the four-point correlation function~(\ref{eq:four:point}) involves only two distinct points $x_1=0$
and $x_2=z$. As a result, the expression~(\ref{eq:four:point}) can be factorized into two bilocal correlation
functions with the same Schwinger lines. In the axial gauge, $z_\mu A_\mu(z) = 0$, the
Schwinger line~(\ref{eq:Phi}) is equal to unity, and therefore one arrives to the identification $G \equiv F$,
which is valid only in this gauge. However, since both the topological density~(\ref{eq:Q}) and the bilocal
correlator~(\ref{eq:G}) are the gauge-invariant quantities, our results -- obtained with in the axial gauge --
must be gauge--independent.

The correlation function of two topological densities~(\ref{eq:Q}) is calculated with the help of
Eqs.~(\ref{eq:g2D}) and~(\ref{eq:four:point}):
\begin{eqnarray}
\langle q(0) q(z) \rangle & = & \frac{3}{32 \pi^4} \frac{N^2}{N^2-1} \Bigl[D(z)+D_1(z)\Bigr] \nonumber \\
& & \hspace{-7mm} \times \Bigl\{2 \Bigl[D(z)+D_1(z)\Bigr] + z_\mu \frac{\partial D_1(z)}{\partial z_\mu}\Bigr\}\,.
\label{eq:qq}
\end{eqnarray}
In a physical language the use of the factorization prescription~(\ref{eq:four:point}) in the pseudoscalar channel
implies that we have associated (a tower of) pseudoscalar glueballs -- which mediate the interaction between the
pseudoscalar $q$-probes -- by two non-perturbatively dressed gluons. For a more appropriate treatment of the
glueballs in the field correlator formalism see Ref.~\cite{ref:glueball:Simonov}.

It is important to notice that the two-point correlation function~(\ref{eq:qq}) in Euclidean
space-time must always be negative for non-zero $z$,
\begin{eqnarray}
\langle q(0) q(z) \rangle < 0 \quad \mbox{for}\quad |z|>0\,.
\label{eq:negativity}
\end{eqnarray}
due to a reflection positivity property and a pseudoscalar nature of the topological charge~\cite{ref:negativity}.
Using the parametrization for the bilocal correlators~(\ref{eq:D:D1}) as well as
the fitting results obtained with the help of the lattice
simulations~\cite{ref:Adriano:pure,ref:Adriano:QCD,ref:new:fits}
one can immediately check that the negativity requirement is indeed satisfied for large
enough distances (typically, for $|z| \gtrsim 1.5 \, \mathrm{fm}$). However, at smaller
distances the correlator (\ref{eq:qq}) becomes positive due to large positive contributions coming
from the perturbative parts of the functions $D(z)$ and $D_1(z)$. These parts are given
in Eq.~(\ref{eq:D:D1}) by the terms proportional to the factors $b_0$ and $b_1$.

It is important to stress that the positivity of the correlator~(\ref{eq:qq})
(calculated with the parameters taken from the lattice data~\cite{ref:Adriano:pure,ref:Adriano:QCD,ref:new:fits})
has an apparent inconsistency with the asymptotic freedom of the Yang-Mills theory. In fact, at small distances the physics must be
dominated by the perturbative corrections. For example, in the tree
order (here $\alpha_s = g^2/(4 \pi)$) one has
\begin{eqnarray}
D^{\mathrm{tree}}(z) = 0\,, \qquad D^{\mathrm{tree}}_1(z) =  \frac{2 \alpha_s}{\pi} \frac{N^2-1}{N} \frac{1}{|z|^4}\,,
\end{eqnarray}
which leads to the known negative-valued result,
\begin{eqnarray}
{\langle q(0) q(z) \rangle}^{\mathrm{tree}} = - \frac{3 \alpha_s^2}{4 \pi^6} \frac{N^2 - 1}{{|z|}^8} < 0\,.
\end{eqnarray}
One can check that perturbative corrections to $D(z)$ and $D_1(z)$
structure functions~\cite{ref:corrections} do not spoil this result.
Note that we do not discuss here
non-perturbative physics at short distances since the corresponding terms must anyway be
small compared to the leading perturbative result.

Moreover, in order for the correlator~(\ref{eq:negativity}) to be negative at small distances,
the parameters $b_0$ and $b_1$ of the parametrization~(\ref{eq:D:D1}) must satisfy the
condition
\begin{eqnarray}
b_0 < b_1\,.
\label{eq:bb}
\end{eqnarray}
However, neither of available results of the lattice
simulations~\cite{ref:Adriano:pure,ref:Adriano:QCD,ref:new:fits}
of the pure Yang-Mills theory and QCD agrees with the requirement~(\ref{eq:bb}).

The observed inconsistency with the perturbative part of the lattice results
and the reflection positivity does not however undermine
non-perturbative calculations utilizing the field correlator method.
A plausible explanation of the contradiction may be related to the cooling procedure used
in the lattice simulations. In fact, the cooling may
affect the short range part of the bilocal correlators, leading to a modification of the
coefficients $b_0$ and $b_1$ as compared with their values in the uncooled vacuum. In Ref.~\cite{ref:Adriano:pure}
it is clearly demonstrated that the field correlator at a fixed short separation,
e.g. $|z| \approx 0.3 \, \mbox{fm}$, in the course of the cooling
increases by a factor of 3 from its initial negative value to a positive value at a plateau.
Moreover, the fits of the numerical data for the correlators were done at higher separations $z$,
and therefore we do not expect that the
fitting results represent correctly the short distance physics. As a consequence, there is no
contradiction between the lattice results~\cite{ref:Adriano:pure,ref:Adriano:QCD,ref:new:fits}
and the reflection positivity~(\ref{eq:negativity}).

Since the aim of this paper is to evaluate the non-perturbative contribution to the susceptibility of the topological charge,
our results should not be affected by uncertainties in determination of the perturbative part of the bilocal field correlators.

Another important remark is that in our calculations given below we neglect the contact term~\cite{ref:negativity} in the
correlator of the local topological densities~(\ref{eq:qq}). This term should appear as a singular $\delta$--like function
at zero separations, $z=0$, in order to reconcile the apparent positivity of the susceptibility~(\ref{eq:Q:susc}) with the
negativity requirement~(\ref{eq:negativity}).
Unfortunately, the contact term cannot be accounted in our calculations because we are constrained
(by results of the lattice simulations) to use the particular prescription~(\ref{eq:D:D1}) of the field correlators~(\ref{eq:g2D}).
Moreover, we expect that in the quoted lattice calculations this term cannot be accessible anyway because these
numerical calculations use the cooling method which must destroy (or affect substantially) the short distance physics.
On the other hand, we know that the (soft) cooling does not affect the chiral and bulk topological properties of the
QCD vacuum~\cite{ref:Negele}. In terms of the correlation function~(\ref{eq:qq})
the latter observation may indicate that in the course of the cooling the non-perturbative part of the zero-distance singularity
is not destroyed literally but it is rather shifted towards longer distances. If true, this means that we can neglect the $z=0$
singularity provided we use only the non-perturbative $z \neq 0$ part which is obtained with the use of the cooling method.
With this justification in mind we now continue with explicit calculations.

We use the data of Ref.~\cite{ref:new:fits} where the long-range tail -- left intact by the cooling procedure
-- was fitted only by the non-perturbative terms of Eqs.~(\ref{eq:D:D1}). The fits (with vanishing perturbative part,
$b_0=b_1=0$) provide us with the following results~\cite{ref:new:fits}
\begin{eqnarray}
D^{\NP}(0) & \equiv & A_0    = 3.62(19) \, \Lambda_L^4 = 0.212(11)\,\mbox{GeV}^4\,, \nonumber\\
D^{\NP}_1(0) & \equiv & A_1  = 1.23(7) \, \Lambda_L^4 = 0.072(4) \,\mbox{GeV}^4\,, \qquad
\label{eq:data}\\
& & \hspace{-15.5mm} T_g = \frac{1}{183(3) \, \Lambda_L} = \frac{1}{0.900(14)\,\mbox{GeV}} = 0.222(4)\,\mbox{fm}\,, \nonumber
\end{eqnarray}
written in various units for the sake of convenience. Here the superscript ``NP'' stands for ``non-per\-tur\-ba\-tive'', and
$\Lambda_L = 4.92\,\mbox{MeV}$ denotes a lattice renormalization scale chosen in Ref.~\cite{ref:new:fits}.

Substituting Eq.~(\ref{eq:qq}) into Eq.~(\ref{eq:Q:susc}) and taking non-perturbative parts
of the correlators~(\ref{eq:D:D1}) we arrive to
\begin{eqnarray}
\chi = \frac{9}{64 \pi^2} \frac{N^2}{N^2-1} D^{\NP}(0) \Bigl[D^{\NP}(0) + D^{\NP}_1(0)\Bigr] T_g^4\,.
\label{eq:chi}
\end{eqnarray}

Finally, a direct evaluation of the susceptibility~(\ref{eq:chi}) with the help of the non-perturbative part
of the field correlators gives us:
\begin{eqnarray}
\chi^{1/4}_{\mathrm{theor}} = 196(7) \,\mbox{MeV}\,.
\label{eq:chi:direct:num}
\end{eqnarray}
Here we used the parameters~(\ref{eq:data}) from Ref.~\cite{ref:new:fits} as a numerical input.

It is remarkable that our result~(\ref{eq:chi:direct:num}) -- based on the direct evaluation of
the topological susceptibility -- coincides (within the small error bars) with the numerical
value
\begin{eqnarray}
\chi^{1/4}_{\mathrm{lattice}} = 193(9) \, \mbox{MeV}\,.
\label{eq:lattice}
\end{eqnarray}
obtained recently in lattice simulations of pure $SU(3)$ gauge theory~\cite{ref:chi:lattice}.
One should notice, however, that the value of the error in our estimation~(\ref{eq:chi:direct:num}) comes only from
the numerical error of the lattice data~(\ref{eq:data}). Consequently, the error in Eq.~\eq{eq:chi:direct:num} does not reflect any systematic
uncertainty which could be related to our assumptions implied in the course of the derivation of Eq.~(\ref{eq:chi}).

In conclusion, we stress that in our calculation of the topological susceptibility
we have used the particular prescriptions related to
(i) the renormalization of the ultraviolet divergences (self-energy) associated with the Schwinger lines,
(ii) the path-dependence of the field correlation functions, and
(iii) the presence of the divergent contact term in the correlators of the topological charge densities.
After a proper treatment of these subtle issues we arrive to the analytical result for the topological susceptibility~(\ref{eq:chi}).
In the case of $SU(3)$ gauge theory we get the numerical value~(\ref{eq:chi:direct:num}) which is in a very good agreement
with the recent lattice estimation~(\ref{eq:lattice}).

\vskip 3mm

The authors are grateful to Yu.A.Simonov for
a critical reading of the manuscript and for many useful suggestions.
The work is supported by the grants RFBR  05-02-16306a, RFBR-DFG 06-02-04010
and by a STINT Institutional grant IG2004-2 025.
M.N.Ch. is grateful to the members of the Department of Theoretical
Physics at Uppsala University for the kind hospitality
and stimulating environment.

\end{document}